\pdfoutput=1 		
\newif\ifextended
\extendedtrue 	 	
\ifextended
\documentclass[twocolumn]{autart_modPreprint}	
\else
\documentclass[twocolumn]{autart}    
\fi

\ifextended
\fi

\usepackage{graphicx}          

\usepackage{epstopdf}	
\usepackage[cmex10]{amsmath}
\usepackage{amsfonts,amssymb,bm}
\usepackage{color}
\usepackage{algorithm}
\floatname{algorithm}{\small Algorithm}

\usepackage{graphicx,color}			
\usepackage{url}

\usepackage{natbib} 	

\graphicspath{{.},{figures/}}
\newcommand{\deltac}{\delta^\text{\normalfont{ctrl}}}
\newcommand{\deltae}{\delta^\text{est}}

\newcommand{\transpose}{\text{T}}
\DeclareMathOperator*{\avg}{avg}
\DeclareMathOperator*{\diag}{diag}

\newcommand{\field}[1]{\mathbb{#1}}
\newcommand{\R}{\field{R}}
\newcommand{\N}{\field{N}}

\newcommand{\ie}{i\/.\/e\/.\/~}
\newcommand{\eg}{e\/.\/g\/.\/~}
\newcommand{\cf}{cf\/.\/~}
\newcommand{\fig}{Fig\/.\/~}
\newcommand{\tab}{Table~}
\newcommand{\sect}{Section~}

\newcommand{\algo}{Algorithm~}

\newcommand{\The}{Theorem~}
\newcommand{\Lem}{Lemma~}

\newcommand{\Assump}{Assumption~}

\newcommand{\Rema}{Remark~}

\newlength{\spaceMatRow}
\usepackage{fancyhdr}

\newcommand{\mytitle}{\textbf{Technical Report.}
This technical report served as the basis for the paper titled ``Event-based State Estimation: An Emulation-based Approach,'' which was submitted and accepted at \emph{IET Control Theory \& Application} and is subject to Institution of Engineering and Technology Copyright. When the final version is published, the copy of record will be available at the IET Digital Library. (doi: 10.1049/iet-cta.2016.1021)
}

\begin{document}
\thispagestyle{fancy}	

\begin{frontmatter}

\title{Distributed Event-based State Estimation\thanksref{footnoteinfo}\vspace{-0.5cm}} 

\thanks[footnoteinfo]{This work was supported in part by the Swiss National Science Foundation, the Max Planck Society, and the Max Planck ETH Center for Learning Systems.
This report is partially based on preliminary results in \citep{Tr12,Tr14}.
}

\author[MPI]{Sebastian Trimpe}\ead{strimpe@tuebingen.mpg.de}    

\address[MPI]{Max Planck Institute for Intelligent Systems, Autonomous Motion Department, 72076 T\"ubingen, Germany}  

\begin{keyword}
Event-based estimation, distributed estimation, state observers, networked control systems.
\vspace{-1mm}
\end{keyword}

\begin{abstract}                          
An event-based state estimation approach for reducing communication in a networked control system is proposed.  Multiple distributed sensor-actuator-agents observe a dynamic process and sporadically exchange their measurements and inputs over a bus network.  Based on these data, each agent estimates the full state of the dynamic system, which may exhibit arbitrary inter-agent couplings. Local event-based protocols ensure that data is transmitted only when necessary to meet a desired estimation accuracy.  
This event-based scheme is shown to mimic a centralized Luenberger observer design up to guaranteed bounds, and stability is proven in the sense of bounded estimation errors for bounded disturbances.
The stability result extends to the distributed control system that results when the local state estimates are used for distributed feedback control.  Simulation results highlight the benefit of the event-based approach over classical periodic ones in reducing communication requirements.
\vspace{-4mm}
\end{abstract}

\end{frontmatter}

\thispagestyle{fancy}	

\section{Introduction}
\label{sec:intro}
In almost all control systems today,
data is processed and transferred between the system's components periodically.
%
While periodic system design is often convenient and well understood \citep{AsWi97}, it involves an inherent limitation: data is processed and transmitted at  predetermined time instants, irrespective of the current state of the system or the information content of the data. That is, system resources  
are used regardless of whether there is any need for processing and communication or not. This becomes prohibitive when resources are scarce, such as in networked or cyber-physical systems \citep{HeNaXu07,KyKu12} where multiple controllers share a communication medium, or wireless sensor networks \citep{AkSuSaCa02} where communication is typically a main consumer of battery power. 
%

Because of the limitations of traditional design methodologies for resource-constrained problems, aperiodic or event-based 
strategies have recently received a lot of attention in the controls community (see e.g.\
\citet{Le11,HeJoTa12,GrHiJuEtAl14,Ca14}), 
as well as related disciplines (\citet{DeLiCuPo10} and \citet{Ts10}, for example).
With event-based methods, data is transmitted or processed only when certain \emph{events} indicate that an update is required, for example, to meet some control or estimation specification.  Thus, resources are used \emph{only when required} and saved otherwise.

\begin{figure}[tb]
\centering
\includegraphics[scale=.9]{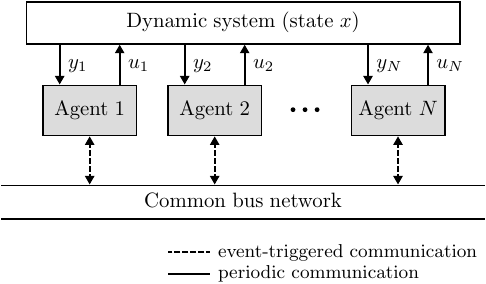}
\caption{Networked control system. Multiple distributed agents make observations $y_i$ of a dynamic system and apply control actions $u_i$.  The agents communicate via a common bus. The development of an event-based scheme for reducing inter-agent communication is the object of this report.
}
\label{fig:networkArchitecture}
\end{figure}

In this report, we propose a novel event-based scheme for distributed  state estimation. We consider the networked control system (NCS) shown in \fig \ref{fig:networkArchitecture}, where multiple sensor-actuator-agents observe and control a dynamic system, and exchange data with each other over a common bus.  Each agent shall estimate the full state of the dynamic system from its local sensors and data received over the bus. Because no local observability is assumed, agents rely on the inter-agent communication to solve the estimation problem.  However, in order to limit network traffic,
local event triggers on each agent ensure that updates are sent only when needed. 
%
%
The local state estimates can be used for feedback control to compute commands to the local actuators.

While the method can thus be used both for distributed event-based \emph{state estimation} and \emph{control}, the primary focus herein is on the state estimation problem.  Since the developed event-based protocols for measurement and input communication serve the purpose of obtaining bounded state estimates, we refer to this method as \emph{event-based state estimation}. The proposed method is \emph{distributed} in the sense that data from distributed sensors is required for stable state estimation, and that transmit decisions are made locally on each agent. 

The common bus (\cf \fig \ref{fig:networkArchitecture}) is a key component of the developed event-based approach.  When one agent transmits sensor and input data, these can be received by all other agents.  
This will allows the agents to compute consistent estimates and use these for the triggering decisions (while inconsistencies can still happen due to data loss  or delay).
Wired
%
fieldbus systems, which support broadcast communication between multiple devices, are common in industry \citep{Th05}.
%
%
%
%
%
%
Recently, \citet{FeZiMoTh12} 
have shown that common bus concepts are feasible even for low-power wireless networks.

The approach for distributed event-based estimation proposed herein mimics a classic centralized, periodic, linear state observer design. The gains of the distributed state observers\footnote{The terms \emph{state observer} and \emph{state estimator} are used synonymously in this report.} are simply taken from the centralized design. In each step, the appropriate subset of gains is chosen corresponding to which measurements have been received. 
The main focus of this report is to prove stability of the resulting switching and distributed estimation scheme. Challenges are 1) the switching nature of the observers \citep{BoLu02}, and 2) the distributed arrangement of estimators coupled through the dynamics and the inter-agent communication.  
The stability results for state estimation extend in a straightforward manner to the control system resulting when local estimates are used for feedback.
Beyond stability of the centralized design, no other assumptions are required.

%

Preliminary results of those herein were presented in the conference papers \citep{Tr12,Tr14}. Compared to these, the framework is extended herein to also include event-based communication of control inputs, which further reduces network traffic. In addition, we streamlined the presentation, significantly expanded review of related work, and added new simulation examples.

\subsection{Related work}

Early work on event-based state estimation concerned problems with a single sensor and estimator node, \eg a sensor transmitting local state estimates to a remote estimator.  \citet{ImBa05} and \citet{RaMoBa06} seek optimal sending rules for finite-horizon problems when the sensor has a fixed number of allowed transmissions, while \citet{XuHe04} consider infinite-horizon problems.  Typically, the optimal sending strategies are found to be policies with time-varying thresholds for finite-horizon problem, and constant thresholds for infinite-horizon problems, \citep[p.~340]{Le11}. Because we are primarily interested in long-term behavior (stability) in this work, and also for simplicity of implementation, we consider constant thresholds.

Different types of stationary triggering policies have been suggested in literature.  With the \emph{send-on-delta} (SoD) protocol \citep{Mi06}, transmissions are triggered based on the difference of the current and last-transmitted measurement.   While SoD and variations thereof
%
can be considered as general-purpose triggers, a number of triggering mechanisms have been tailored to the state estimation problem.  \emph{Measurement-based triggering} \citep{TrDAn11} or \emph{predicted sampling} \citep{SiKeNo14} places a threshold on the measurement innovation; that is, the difference of the current measurement and its prediction based on a process model.  
%
%
Thus, this trigger captures a quantity directly relevant for the state estimation.  
\citet{WuJiJoSh13} use the same trigger, but apply a transformation to decorrelate the innovation. Considering the variance of the innovation instead yields \emph{variance-based triggering} \citep{TrDAn11c,TrDAn14b}.
%
\citet{MaSi10} proposed \emph{relevant sampling}, where the relative entropy of prior and posterior state distribution is employed as a measure of information gain.  While all these triggers are deterministic, \citet{HaMoWuWeSiSh15} have suggested a stochastic trigger, which maintains a Gaussian innovation process and thus facilitates theoretical analysis. In this work, we use measurement-based triggers, which are straightforward to implement and effective for event-based estimation (see comparisons by \cite{SiKeNo14,TrCa15}).

Besides the event trigger, the second main component of an event-based estimation method is the estimation or filtering algorithm.  In the context of event-based estimation, a particular focus is on how the information contained in \emph{non-events}, \ie instants when no data is transmitted, is handled.  For example, not receiving an update with SoD means that the current measurement is within a range of the last transmitted one. Taking a probabilistic viewpoint, it is clear how to fuse this information: the Bayesian estimator simply conditions on the available information, whether it is a data point or a set.  However, this typically yields a non-Gaussian posterior and intractable algorithms. Therefore, different approximations have been suggested in literature \citep{HeAs2006,SiLa12,SiNoHa13}.
%
%
If one ignores the extra information from non-events in favor of a standard implementation, a time-varying Kalman filter (KF) can be used (\cite{TrDAn11}, for example).  Herein, we use the same linear estimator structure as for the standard KF, but with constant, switching gains taken from a centralized design.  The approach thus has the lowest computational complexity of all mentioned algorithms.

Most of the above works focus on single-link estimation problems. To the best of the author's knowledge, \emph{distributed} event-based state estimation with multiple sensor and estimator nodes, and dynamic couplings was first studied by \citet{TrDAn11}.  Considering the same architecture as in \fig \ref{fig:networkArchitecture}, each agent implements KFs as  estimators and measurement-based triggers for transmit decisions.   While \citet{YoTiSo02} had proposed the use of state estimators for the purpose of saving communication before, they do not employ state estimation in the usual sense.  Instead of \emph{fusing} model-based predictions with incoming data, they \emph{reset} parts of the state vector.
%
%
%
%
Later results on distributed event-based estimation 
for related, but different scenarios to the one herein, include the work by \cite{WeArJo12, BaBeCh12, ShChSh14c}. These works consider a central fusion node, while estimation is distributed herein.
\citet{YaZhZhYa14,LiWaHeZh15} treat problems where estimation is distributed over several nodes communicating according to a graph topology.
%
However, these works employ simpler SoD-type triggers, which do not take advantage of model-based predictions for making more effective transmit decisions (see \citep{TrCa15}).  None of the mentioned references treats the problem of emulating a given centralized observer design with a distributed and event-triggered implementation, which does not rely on any assumptions except stability of the centralized design.

When the event-based state estimators are connected to state-feedback controllers, this structure represents a \emph{distributed event-based output-feedback control} system.  
\citet{WaLe08,WaLe11} and \citet{MaTa08,MaTa11} were among the first to discuss distributed or decentralized implementations of event-based control for physically coupled systems.  Whereas \citeauthor{WaLe08} assume weakly coupled dynamic systems, \citeauthor{MaTa08} have no assumption on the system coupling, yet rely on a centralized controller.  While we use distributed control computations and distributed event triggering without any assumption on the system couplings, we rely on a common bus network.
In contrast to the mentioned works, which assume perfect state information, we deal with noisy output measurements.
Other results on distributed or decentralized event-based control for different scenarios include \citep{DoHe12,DiFrJo12,StVeLu13,PeSaWi13,MoHi14,TaCh14,SiStGrLu15}, for example.
%


The key ideas of the approach to event-based \emph{estimation} taken herein  are conceptually related to the approach for event-based \emph{control} by \cite{LuLe02}.  
Therein, the authors design an event triggering scheme such that the difference between the state of a reference system with continuous feedback and the state of the event-based design is bounded.  Here, a centralized Luenberger observer serves as the reference estimator,
and event triggers are designed such that the event-based implementation mimics the reference design.

\subsection{Notation}
\label{sec:notation}
$\R$, $\N$, and $\N_N$ denote real numbers, positive integers, and the set $\{1, 2, \dots, N\}$, respectively.  
Where convenient, vectors are expressed as tuples $(v_1, v_2, \dots)$, where $v_i$ may be vectors themselves, with dimension and stacking clear from context.
For $v \in \R^n$ and $q \in [1,\infty]$, $\|v\|_q$ (or simply $\|v\|$) denotes the vector H\"older norm
$\|v\|_q = ( \sum\nolimits_{j=1}^n |v_j|^q )^{1/q}$ for $1\leq q < \infty$, and $\|v\|_q = \max_{j \in \{1,\dots,n\}} |v_j|$ for $q=\infty$, \citep{Be05}.
For a matrix $A$, $\|A\|_q$ (or simply $\|A\|$) denotes the matrix norm of $A$ induced by the chosen vector norm.
For a time-indexed sequence $v = \{v(0), v(1), v(2), \dots \}$, $\|v\|_\infty$ denotes the $\ell^\infty$ norm $\|v\|_\infty := \sup\nolimits_{k\geq 0} \, \|v(k)\|$, \citep{CaDe91}.
%
%
%
We use $\avg_{i=1,\dots,N}(v_i) := \frac{1}{N} \sum\nolimits_{i=1}^N v_i$ to denote the average of $v_1, \dots, v_N$ and 
omit the index ``$i=1,\dots,N$'' when clear from context.
For an estimate of $x(k)$ computed from measurement data until time $\ell \leq k$, we write $\hat{x}(k|\ell)$.  For ease of notation, we also write $\hat{x}(k)$ for $\hat{x}(k|k)$.  
Finally, a matrix is called stable if all its eigenvalues have magnitude strictly less than one.

\section{Preliminaries and Problem Formulation}
\label{sec:problemFormulation}
In this section, we formally introduce the considered NCS and the centralized observer-based control design, which serves as a reference for the distributed and event-based scheme developed in this report.

\subsection{Networked control system}
We consider the NCS in \fig \ref{fig:networkArchitecture} with $N$ sensor-actuator-agents.
The dynamics of the entire system are described by 
the discrete-time, linear, time-invariant system
\begin{align}
x(k) &= A x(k-1) + B  u(k-1) + v(k-1) \label{eq:system_x} \\
y(k) &= C  x(k) +w(k) \label{eq:system_y}
\end{align}
with time index $k \in \N$ (corresponding to a sampling time $T_s$), state $x(k) \in \R^n$, control input $u(k) \in \R^{q}$, measurement $y(k) \in \R^{p}$, disturbances $v(k) \in \R^n$, $w(k) \in \R^{p}$, and all matrices of corresponding dimensions.  We assume that $(A,B)$ is stabilizable and $(A,C)$ is detectable.  
No specific assumptions on the characteristics of the disturbances $v(k)$ and $w(k)$ are made; they can be random variables or deterministic disturbances. 
Most results herein apply to both scenarios.

The vectors $u(k)$ and $y(k)$ represent the collective inputs and measurements of the $N$ agents.  We use $y_i(k) \in \R^{p_i}$ and $u_i(k) \in \R^{q_i}$ to denote the individual measurements and inputs by agent $i \in \N_N$; that is, 
$u(k) = ( u_1(k),$ $u_2(k),$ $\dots,$ $u_N(k) )$ and $y(k) = ( y_1(k), y_2(k), \dots, y_N(k) )$.
With this, \eqref{eq:system_x} and \eqref{eq:system_y} can be rewritten as
\begin{align}
x(k) &= A x(k-1) + \sum_{i \in \N_N} B_i u_i(k-1) + v(k-1) \label{eq:system_x_long} \\
y_i(k) &= C_i x(k) +w_i(k)  \qquad \forall \, i \in \N_N  \label{eq:system_yi}
\end{align}
where the dimensions of $B_i$, $C_i$, and $w_i$ follow from $y_i$ and $u_i$.
Agents can be heterogeneous with different types and dimensions of measurements and inputs, including the possibility of no sensors ($p_i = 0$) or actuators ($q_i=0$).
To avoid special treatment of this case in the following, a summation involving a signal of zero dimension (such as \eqref{eq:system_x_long} for some $q_i=0$) is understood to not include the corresponding summand.
We do not make any assumption on stabilizability and detectability for the individual agents; that is, $(A,B_i)$ can possibly be not stabilizable, and $(A,C_i)$ not detectable.

The agents are connected over a common-bus network as shown in \fig \ref{fig:networkArchitecture}, over which they can exchange measurements and inputs.  Supported by the bus, all agents will receive the data if one agent communicates.  Agents are assumed to be synchronized in time, 
and network communication is abstracted as instantaneous.

\subsection{Centralized design with periodic communication}
\label{sec:centralizedController}
For stabilizing \eqref{eq:system_x}, \eqref{eq:system_y}, a centralized control system, which has periodic access to all measurements $y$ and computes all inputs $u$, can readily be designed as the combination of a state estimator and a state-feedback law.
A linear, time-invariant state estimator is given by
\begin{align}
&\hat{x}_\text{c}(k|k-1) = A  \hat{x}_\text{c}(k-1|k-1) + B  u(k-1) \label{eq:FCSE1} \\
&\hat{x}_\text{c}(k|k) = \hat{x}_\text{c}(k|k-1) + L  \big(y(k) - C \, \hat{x}_\text{c}(k|k-1) \big) 
\label{eq:FCSE2}
\end{align}
with static estimator gain $L$.  For the later development, we rewrite \eqref{eq:FCSE2} as
\begin{align}
\hat{x}_\text{c}(k|k) &= \hat{x}_\text{c}(k|k-1) + \!\! \sum_{\ell\in \N_N} \!  L_\ell \big( y_\ell(k) - C_\ell \hat{x}_\text{c}(k|k-1) \big) \nonumber \\
&\text{with} \quad
L = [L_1, L_2, \dots, L_N], \, L_i \in \R^{n \times p_i} .
\label{eq:FCSE2_rewritten} 
\end{align}
%
We further consider the static state-feedback controller 
\begin{equation}
u(k) = F \hat{x}_\text{c}(k)
\label{eq:stateFeedback}
\end{equation}
with controller gain $F$ (recall that $\hat{x}_\text{c}(k) = \hat{x}_\text{c}(k|k)$). 
The centralized closed-loop control system \eqref{eq:system_x}, \eqref{eq:system_y}, \eqref{eq:FCSE1}, \eqref{eq:FCSE2}, and \eqref{eq:stateFeedback} is described by
\begin{align}
x(k) &= (A\!+\!BF) x(k\!-\!1) - BF \epsilon_\text{c}(k\!-\!1) + v(k\!-\!1) \label{eq:closedLoopCentralized_state} \\
\epsilon_\text{c}(k) &= (I\!-\!LC)A \epsilon_\text{c}(k\!-\!1) + (I\!-\!LC)v(k\!-\!1) - Lw(k)
\label{eq:closedLoopCentralized_est}
\end{align}
%
%
where $\epsilon_\text{c}(k) = x(k) - \hat{x}_\text{c}(k)$ is the centralized estimation error.  A standard result and obvious from \eqref{eq:closedLoopCentralized_state}, \eqref{eq:closedLoopCentralized_est}, the closed-loop system is stable if both estimation error and state dynamics are stable, which we shall assume:
%
\begin{assum}
\label{ass:centralEst}
$L$ is such that $(I-LC)A$ is stable.
\end{assum} \vspace{-2ex}
\begin{assum}
\label{ass:centralCtrl}
$F$ is such that $A+BF$ is stable.
\end{assum}
%
%
Because $(A,B)$ is stabilizable and $(A,C)$ detectable, such gains exist and can be computed using standard methods (see \eg \citep{AsWi97}).
\subsection{Problem statement}
\label{sec:problem_subsec}
We seek a distributed and event-based state estimation design that emulates the centralized estimator \eqref{eq:FCSE1} and \eqref{eq:FCSE2} up to guaranteed bounds and with lower average communication requirements. Furthermore, the stability results for estimation shall be extended to the distributed event-based control system that is obtained when the state estimates are used for state-feedback control \eqref{eq:stateFeedback}.
State estimators, controllers, and event triggers are to be implemented locally on all agents of the NCS in \fig \ref{fig:networkArchitecture} without central coordination.

%

\section{Key Ideas: Event-based State Estimation\newline with Perfect Communication}
\label{sec:DEBSE_idealized}

In this section, the key ideas for the event-based estimation approach are developed by first considering an idealized scenario with perfect inter-agent communication.

\subsection{Architecture}
We propose a distributed estimation scheme, where each agent maintains an estimate $\hat{x}_i$ of the entire system state $x$ such that $\hat{x}_i$ mimics the centralized estimate $\hat{x}_\text{c}$, but with reduced  communication between the agents.
Because we are interested in local transmit decisions without any central coordination, the key question when designing event-triggers is:
\emph{How can agent $i$ know whether the other agents are currently in need of its sensor data for solving their estimation problems?
}
We propose an architecture, where each agent keeps track of the state information that is commonly available to all agents in the network. This is the state estimate computed from all data that has been broadcast over the common bus, which all agents have access to.  Based on this common estimate, agent $i$ can make an informed transmit decision: if the prediction of agent $i$'s measurement based on the common data is already good enough, there is no need for agent $i$ to send an update; if the prediction is poor, however, sending an update and thus using the network resource is required.  
%

\begin{figure}[tb]
\centering
\includegraphics[scale=.85]{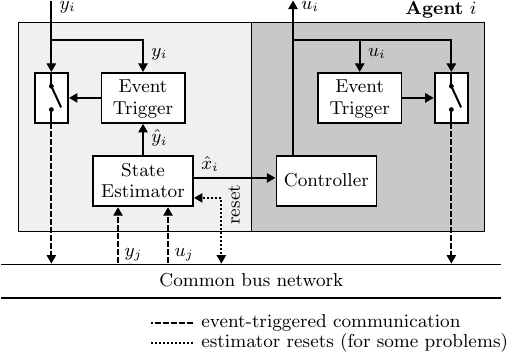}
\caption{Event-based architecture implemented on each agent of the NCS in \fig \ref{fig:networkArchitecture}.  
The light-gray area encompasses the blocks used for state estimation.  The darker area has additional blocks required when local estimates are used for distributed control.
%
%
}
\label{fig:oneAgent}
\end{figure}

The block diagram in \fig \ref{fig:oneAgent} illustrates the components of the proposed event-based estimation architecture as implemented on each agent (light-gray area).  
The \emph{State Estimator} periodically computes an estimate $\hat{x}_i$ of the full system state $x$ based on data received over the bus.  Since all agents read on the bus, the estimates $\hat{x}_i$ are consistent among all agents
and thus represent the common information. 
The \emph{Event Trigger} issues the transmission of the local measurement $y_i$ to all other agents if, and only if, there is a significant mismatch between $y_i$ and its prediction $\hat{y}_i$ from the State Estimator.

%
%
%

\begin{rem}
To ensure consistent estimates among all agents, the State Estimator in \fig \ref{fig:oneAgent} uses the local measurement $y_i$ only when it is also shared over the bus.  
For the purpose of improving local state estimates, one may wish to incorporate $y_i$ in the estimation at every step.
For this purpose,  a second estimator can be introduced that fuses, at every step, the network data \emph{and} the local sensor data (see \cite{TrDAn11}).
\label{rem:ownMeasurement}
\end{rem}

One possible use of local state estimates is for feedback control.  As shown in \fig \ref{fig:oneAgent} (dark-gray area), the estimate is 
fed to a \emph{Controller}, which computes inputs $u_i$ to the local actuator.\footnote{We remark that the use of a Controller is optional.  The method equally applies to pure estimation problems, where $u$ in \eqref{eq:system_x} may be a known reference or nonexistent.} 
With an estimate of the full state at hand, 
each agent can simply implement \eqref{eq:stateFeedback} for its local actuator to emulate the centralized design; namely
\begin{equation}
u_i(k) = F_i \hat{x}_i(k)
\label{eq:stateFeedbackDistributed}
\end{equation}
where $F_i$ is the part of the gain matrix $F$ in \eqref{eq:stateFeedback} corresponding to its local input $u_i$. Since the inputs also need to be available for state estimation, another \emph{Event Trigger} is employed to manage their communication.   

In this section, we shall focus on the event-based estimation problem; that is, the design of the Event Trigger and the State Estimator in the light-gray area in \fig \ref{fig:oneAgent}.  To develop the main ideas and leaving technical details aside, 
we consider a simplified scenario first.

\subsection{Idealizing assumption}

If all estimates $\hat{x}_i(0)$ are initialized identically, and the bus communication is perfect (\ie without any delay and packet drops), all estimates $\hat{x}_i$ will be identical. This, we shall assume throughout this section:
\begin{assum}
\label{ass:identicalEstimates}
$\hat{x}_i(k) = \hat{x}_j(k)$ for all $i,j \in \N_N$, $k \in \N$.
\end{assum}
Clearly, this assumption is unrealistic in practice because it will be violated by a single packet drop or a slight difference in initialization or computation on any two agents.
Thus, we remove it again in \sect \ref{sec:DEBSE}. 

Assuming knowledge of the control gain $F$,
every agent can reconstruct locally the complete control input vector
\begin{equation}
\hat{u}^i(k) = F \hat{x}_i(k)
\label{eq:inputEstByAgenti}
\end{equation} 
and use it for state prediction \eqref{eq:FCSE1}.
Assumption \ref{ass:identicalEstimates} implies that this reconstruction is perfect ($\hat{u}^i(k) = u(k)$) and exchange of inputs between agents is thus not necessary.
Therefore, we focus solely on the event-based communication of measurements in this section.  Communication of inputs and also occasional estimator resets as shown in \fig \ref{fig:oneAgent} shall become relevant later.

\subsection{Event trigger}
\label{sec:EventTriggerMeas}
The Event Trigger on agent $i$ (\fig \ref{fig:oneAgent}, light-gray) decides at every step $k$, whether or not the local measurement $y_i(k)$ is sent to all other agents using the decision rule:
\begin{equation}
\text{transmit $y_i(k)$} 
\; \Leftrightarrow \;
\| y_i(k) - C_i \hat{x}_i(k|k-1) \| \geq \deltae_i 
\label{eq:eventTrigger_MB}
\end{equation}
where $\deltae_i \geq 0$ is a design parameter, $\hat{x}_i(k|k-1)$ is agent $i$'s prediction of the state $x(k)$ based on measurements until time $k-1$ (made precise in the next subsection), 
and $\hat{y}_i(k) = C_i \hat{x}_i(k|k-1)$ is agent $i$'s prediction of its measurement $y_i(k)$.  Hence, $y_i(k)$ is broadcast if, and only if, the prediction deviates by more than the tolerable threshold $\deltae_i$.  
%
%
%
Tuning $\deltae_i$ allows the designer to trade off each sensor's frequency of events (and, hence, the communication rate) for estimation performance.  For later reference, we introduce $\deltae := (\deltae_1, \dots, \deltae_N)$.

We denote by $I(k)$ and $\bar{I}(k)$ the indices of those agents transmitting and not transmitting at time $k$; that is,
\begin{align}
I(k) &:= \{ i \in \N_N \, | \, \| y_i(k) - C_i \hat{x}_i(k|k\!-\!1) \| \geq \deltae_i \}
\label{eq:I} \\
\bar{I}(k) &:= \{ i \in \N_N \, | \, \| y_i(k) - C_i \hat{x}_i(k|k\!-\!1) \| < \deltae_i \} .
\label{eq:Ibar}
\end{align}

\subsection{State estimator}
Agent $i$'s State Estimator (\fig \ref{fig:oneAgent}) recursively computes an estimate $\hat{x}_i(k) = \hat{x}_i(k|k)$ of the system state $x(k)$ from the all measurements $I(1), \dots, I(k)$ transmitted until time $k$.  We propose the following estimator update:
\begin{align}
\hat{x}_i(k|k-1) &= A  \hat{x}_i(k-1|k-1) + B  \hat{u}^i(k-1) \label{eq:EBSE1}  \\
\hat{x}_i(k|k) &= \hat{x}_i(k|k-1) \nonumber \\
&\phantom{=}+ \!\! \sum_{\ell \in I(k)} \!\! L_\ell \big( y_\ell(k) - C_\ell \hat{x}_i(k|k-1) \big)   \label{eq:EBSE2ideal} 
\end{align}
where the observer gains $L_\ell$ are taken from the centralized design \eqref{eq:FCSE2_rewritten}, and
$\hat{u}^i(k-1)$ is agent $i$'s knowledge of the full input vector (here, $\hat{u}^i(k-1) = u(k-1)$ from \eqref{eq:inputEstByAgenti} and \Assump \ref{ass:identicalEstimates}).
Comparing \eqref{eq:EBSE1}, \eqref{eq:EBSE2ideal} to \eqref{eq:FCSE1}, \eqref{eq:FCSE2_rewritten}, we see that the estimator structure is the same, but the event-based estimator updates with a subset $I(k) \subset \N_N$ of all measurements.
If, at time $k$, no measurement is transmitted (\ie $I(k) = \emptyset$), (\ref{eq:EBSE2ideal}) is to be understood such that the summation vanishes; that is, $\hat{x}_i(k|k) = \hat{x}_i(k|k-1)$. In order to ease the presentation, this case is not explicitly mentioned hereafter.

\subsection{Stability analysis}
\label{sec:idealStabAnalysis}
The estimator equations \eqref{eq:EBSE1}, \eqref{eq:EBSE2ideal} and the triggering rule (\ref{eq:eventTrigger_MB}) together constitute the porposed event-based state estimator (EBSE). 
For a fixed sequence $I(1), \dots, I(k)$, the estimator \eqref{eq:EBSE1}, \eqref{eq:EBSE2ideal} is linear.
Because the index set $I(k)$ depends on $y(k)$ by (\ref{eq:I}), however, the EBSE is a nonlinear, switching observer, whose switching modes are governed by the event trigger \eqref{eq:eventTrigger_MB}.
%
%
Stability of the EBSE is discussed next.

Since the main objective is to mimic the centralized reference estimator $\hat{x}_c(k)$ (\cf \sect \ref{sec:problem_subsec}), we first consider the difference $e_i(k) = \hat{x}_c(k) - \hat{x}_i(k)$.
Using \eqref{eq:FCSE1}, \eqref{eq:FCSE2_rewritten}, \eqref{eq:EBSE1}, and \eqref{eq:EBSE2ideal}, straightforward manipulation yields
\begin{align}
e_i(k) 
&=A e_i(k\!-\!1) 
+ \sum\nolimits_{\ell \in \N_N}  L_\ell \big( y_\ell(k) - C_\ell \hat{x}_\text{c}(k|k\!-\!1) \big) \nonumber \\
&\phantom{=}-\!\!\underbrace{\sum\nolimits_{\ell \in I(k)} \!\! L_\ell \big( y_\ell(k) - C_\ell \hat{x}_i(k|k-1) \big)}_{
\sum_{\ell \in \N_N} \! L_\ell (\, \ldots \,) \,\,
- \,\, \sum_{\ell \in \bar{I}(k)} \! L_\ell (\, \ldots \,)
} \nonumber  \\
%
&=(I-LC)A e_i(k-1) \nonumber \\
&\phantom{=}+\!\! \sum\nolimits_{\ell \in \bar{I}(k)} \!\! L_\ell \big( y_\ell(k) - C_\ell \hat{x}_i(k|k-1) \big) 
\label{eq:epsilon_ci}
\end{align}
The error $e_i(k)$ is governed by the stable centralized estimator dynamics $(I-LC)A$ 
with an extra input term, 
which is essentially bounded by the choice of the event-trigger (\ref{eq:eventTrigger_MB}), as can be seen from (\ref{eq:Ibar}) with $\hat{x}_i(k|k-1) = \hat{x}_\ell(k|k-1)$.
We thus have the following result:
\begin{thm}
\label{thm:epsilon_ci_idealized}
Under Assumptions \ref{ass:centralEst} and \ref{ass:identicalEstimates}, the difference $e_i(k)$ between the centralized estimator and the EBSE \eqref{eq:EBSE1}, \eqref{eq:EBSE2ideal}, and \eqref{eq:eventTrigger_MB} is bounded for 
all initial conditions $\hat{x}_i(0)$ and $x_\text{\normalfont{c}}(0)$.  In particular, 
there exist constants $c>0$ and $\rho \in [0,1)$ such that
\begin{equation}
\| e_i \| \leq c \| e_i(0) \| +  \frac{c}{1-\rho} \|L\| \|\delta^\text{\normalfont{est}}\|.
\label{eq:thm_epsilon_ci_idealized}
\end{equation}
\end{thm}
\begin{pf}
With \Assump \ref{ass:identicalEstimates}, \eqref{eq:epsilon_ci} is rewritten as 
\begin{equation}
e_i(k) = (I-LC)A e_i(k-1) + L_{\bar{I}(k)} \Delta_{\bar{I}(k)}(k)
\label{eq:epsilon_ci_rewritten}
\end{equation}
where $\Delta_i(k):=y_i(k)-C_i \hat{x}_i(k|k-1)$, $\Delta_{\bar{I}(k)}(k)$ denotes the vector obtained from  stacking $\Delta_i(k)$, $i \in \bar{I}(k)$, from top to bottom; and $L_{\bar{I}(k)}$ is obtained from stacking $L_i$, $i \in \bar{I}(k)$, from left to right.

By \Assump \ref{ass:centralEst}, $e_i(k) = (I-LC)A e_i(k-1)$ is exponentially stable, and there exist $c>0$ and $\rho \in [0,1)$ such that 
$\| ((I-LC)A)^{k} \| \leq c \rho^{k}$ for all $k \in \N$ \citep[p.~212--213]{CaDe91}.
%
%
%
For $1 \leq q < \infty$, we find
\begin{align}
\|\Delta_{\bar{I}(k)}(k) \|_q^q
= \!\! \sum_{i \in \bar{I}(k)} \!\! \| \Delta_i(k)\|_q^q
\underset{\eqref{eq:Ibar}}{<} \!\! \sum_{i \in \bar{I}(k)} \!\! (\deltae_i)^q
\leq 
\|\deltae \|_q^q , 
\label{eq:proof_boundedness_normq}
\end{align}
and, for $q = \infty$,
\begin{align}
\| \Delta_{\bar{I}(k)}(k) \|_q
= \! \max_{i \in \bar{I}(k)} \|\Delta_i(k) \|_q
\! \underset{\eqref{eq:Ibar}}{<} \! \max_{i \in \bar{I}(k)} \deltae_i 
\leq 
\|\deltae \|_q . 
\label{eq:proof_boundedness_normmax}
\end{align}
Hence, 
$\|\Delta_{\bar{I}(k)}(k) \| < \|\deltae \|$.
Since also $\| L_{\bar{I}(k)} \| \leq \|L\|$, 
the input term $L_{\bar{I}(k)} \Delta_{\bar{I}(k)}(k)$ in (\ref{eq:epsilon_ci_rewritten}) is bounded.  Using these results and applying 
\cite[p.~218, Thm. 75]{CaDe91} yields the inequality \eqref{eq:thm_epsilon_ci_idealized}.\hfill \hfill \qed
\end{pf}

The first term in \eqref{eq:thm_epsilon_ci_idealized}, $c \| e_i(0) \|$, is due to possibly different initial conditions.  
As initial conditions will decay exponentially, $\frac{c}{1-\rho} \|L\| \|\deltae\|$ represents the asymptotic bound.  Choosing $\deltae_i$ small enough, $e_i(k)$ can hence be made arbitrarily small as $k \to \infty$,
and, for $\deltae_i = 0$,
the performance of the centralized estimator is recovered.

The bound (\ref{eq:thm_epsilon_ci_idealized}) holds irrespective of the representation of the disturbances $v$ and $w$ in \eqref{eq:system_x}, \eqref{eq:system_y}.  In particular, it also holds for the case where the disturbances are unbounded, such as for Gaussian noise.  The reason is that \The \ref{thm:epsilon_ci_idealized} concerns $e_i(k) = \hat{x}_c(k) - \hat{x}_i(k)$, the difference between an agent's estimate to the (hypothetical) centralized estimator, and the noise will affect both estimates in the same way.

The actual estimation error $\epsilon_i$ of agent $i$ is
\begin{align}
\epsilon_i(k) &:= x(k) - \hat{x}_i(k)
= \epsilon_\text{c}(k) + e_i(k) . 
\label{eq:estError_ei}
\end{align}
\The \ref{thm:epsilon_ci_idealized} can thus be used to deduce properties of the estimation error $\epsilon_i$ from properties of the centralized estimator.
For example, in a deterministic setting with bounded disturbances $v$, $w$ and thus bounded $\epsilon_\text{c}$, a bound on the estimation error can readily be computed from \eqref{eq:estError_ei} and \The \ref{thm:epsilon_ci_idealized}.  For a stochastic scenario, where $v$, $w$ are zero-mean with known variance, the variance of $\epsilon_\text{c}$ can be determined using standard methods, and  \eqref{eq:estError_ei} and \The \ref{thm:epsilon_ci_idealized} can then be used to compute a bound on the variance of $\epsilon_i$.


\ifextended 
\subsection{Remote estimation scenario}
While \Assump \ref{ass:identicalEstimates} is unrealistic for a multi-agent scenario, it is trivially satisfied when there is only one agent.  An example scenario is shown in \fig \ref{fig:remoteEstimationSingleAgent}.  Therein, a sensor agent observes a dynamic process and sporadically transmits measurements to a remote agent, which computes state estimates for monitoring purposes, for example.  The remote estimator thus implements the state estimator \eqref{eq:EBSE1}, \eqref{eq:EBSE2ideal}.  The sensor node also implements \eqref{eq:EBSE1}, \eqref{eq:EBSE2ideal}, as well as the event trigger \eqref{eq:eventTrigger_MB} for making the transmit decisions.

\begin{figure}[tb]
\centering
\includegraphics[width=0.97\columnwidth]{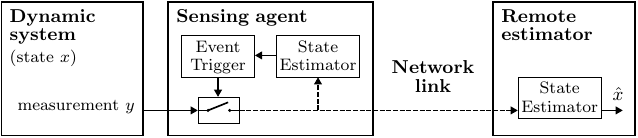}
\caption{Remote estimation scenario with a single sensor.  
}
\label{fig:remoteEstimationSingleAgent}
\end{figure}

The stability analysis of this section directly applies to the stability of the estimator on the sensor node.  If communication from sensor to remote estimator is perfect, the remote estimate is the same as the estimate at the sensor.  However, even if the communication is disturbed, this does not impair stability.  While the disturbance will clearly affect the remote estimate, this one, in turn, has no effect on the estimate at the sensor and thus future transmit decisions. This is fundamentally different for the distributed arrangement as in \fig \ref{fig:networkArchitecture}, where each of the agents computes estimates and makes transmit decisions based on these.  Agents thus influence each other, and stability can actually be lost if two agents' estimates differ, as is discussed in the next section.

\fi		

\section{Distributed Event-based State Estimation}
\label{sec:DEBSE}
In this section, we take up the main ideas from the previous section,
but address two aspects relevant for a distributed implementation in practice.
First, we remove the unrealistic \Assump \ref{ass:identicalEstimates} and analyze stability when agents' estimates can differ ($\hat{x}_i(k) \not= \hat{x}_j(k)$).  
%
Second, we treat the communication of the inputs $u$. 
%

\subsection{Event-based communication of inputs}
Knowledge of the control input $u$ is required in the estimator update \eqref{eq:EBSE1}.  In the previous section, this was trivially achieved since, by \Assump \ref{ass:identicalEstimates}, every agent can compute every other agent's input \emph{exactly} from its local estimate. 
If this is not the case, one has different options.  First, all agents could communicate their input \eqref{eq:stateFeedbackDistributed} periodically over the bus. While this increases network load, it can be a viable option if the number of inputs is comparably small.  
Second, each agent could estimate the full input vector according to \eqref{eq:inputEstByAgenti}; without \Assump \ref{ass:identicalEstimates}, this only yields an approximation of the true input ($\hat{u}^i(k) \approx u(k)$).  It can be shown that the separation principle does not hold for this information pattern even for periodic communication; 
that is, Assumptions \ref{ass:centralEst} and \ref{ass:centralCtrl} are not sufficient for stability.
Notwithstanding, it is possible to specifically design observer gains to achieve closed-loop stability, \citep{MuTr15}.  Since herein, we seek to mimic the centralized observer \eqref{eq:FCSE1}, \eqref{eq:FCSE2} without changing its gain $L$, we opt for a third alternative, where inputs are shared over the common bus, but according to an event-based protocol.

The Controller in \fig \ref{fig:oneAgent} implements \eqref{eq:stateFeedbackDistributed} to compute the local input $u_i$.  
The corresponding Event Trigger uses the following rule to decide when to broadcast $u_i$:
\begin{equation}
\text{transmit $u_i(k)$} \;\; 
\Leftrightarrow \;\;
\|u_i(k) -  u_{i, \text{last}}(k) \| \geq \deltac_i
\label{eq:eventTriggerCtrl}
\end{equation}
where $\deltac_i \geq 0$ is a tuning parameter, and $u_{i, \text{last}}(k)$ is the last input that was sent. That is, we employ a simple send-on-delta scheme for input communication. 
Analogous to \sect \ref{sec:EventTriggerMeas},
we define 
\begin{align}
I^\text{ctrl}(k) &:= 
\big\{ i \in \N_N \, | \, \|u_i(k) -  u_{i, \text{last}}(k) \| \geq \deltac_i \big\} 
\end{align}
and $\deltac := (\deltac_1, \dots, \deltac_N)$.

Each agent keeps track of the last known inputs of all other agents. Agent $i$'s estimate of agent $j$'s input is 
\begin{equation}
\hat{u}^i_j(k)
=
\begin{cases}
u_j(k) & \text{if \eqref{eq:eventTriggerCtrl} triggered} \\
\hat{u}^i_j(k-1) & \text{otherwise}.
\end{cases}
\label{eq:inputEstSingle}
\end{equation}
Accordingly, all agents use $\hat{u}^i(k-1) = ( \hat{u}^i_1(k-1), \hat{u}^i_2(k-1), \dots, \hat{u}_N(k-1) )$ 
as estimate of the full input vector for the estimator update \eqref{eq:EBSE1}.  Since all agents use the same input vector estimate, we write $\hat{u}(k)=\hat{u}^i(k)$ for simplicity.
For the input estimation error $\tilde{u}(k) := u(k) - \hat{u}(k)$, the following is immediate:
%
\begin{lem}
$\| \tilde{u} \|_\infty \leq \| \deltac \|$.
\label{lem:inputErrorsBounded}
\end{lem}
%

\begin{rem}
Each agent uses the last communicated value $\hat{u}_i(k-1)$ of its own input $u_i(k-1)$ in the estimator update, even though the true input $u_i(k-1)$ could be used instead.  This is analogous to not always using the latest local measurement in the estimator (\cf \Rema \ref{rem:ownMeasurement}) in order to keep consistent estimates.
However, the true local input can also be used in the estimator update with only slight modification of the following derivations.
\end{rem}

\subsection{Inter-agent differences}
\label{sec:interAgentDiff}
To account for differences in the agents' estimates (\ie removing Assumption \ref{ass:identicalEstimates}),
we introduce a disturbance signal $d_i$ in each estimate and \emph{replace} \eqref{eq:EBSE2ideal} with
\begin{align}
\hat{x}_i(k|k) &= \hat{x}_i(k|k\!-\!1) \! + \! \sum\nolimits_{\ell \in I(k)} \!\!\!\!\! L_\ell \big( y_\ell(k) - C_\ell \hat{x}_i(k|k\!-\!1) \big) \nonumber \\
&\phantom{=} +d_i(k).  \label{eq:EBSE2}
\end{align}
The disturbance $d_i$ may stem from, for example, unequal initialization, different computation accuracy, or imperfect communication of measurements or inputs.

\begin{rem}
\label{rem:packetDrops}
We emphasize that the disturbance $d_i$ is introduced in \eqref{eq:EBSE2} solely for the purpose of analysis. When implementing the estimator, \eqref{eq:EBSE2ideal} is used in the sense that all measurements received by agent $i$ at time $k$ are included in the summation.  
Packet drops can be represented through $d_i$ as follows: if $y_\ell(k)$, $\ell \in I(k)$ is a measurement not received at agent $i$, then $d_i(k) = -L_\ell ( y_\ell(k) - C_\ell \hat{x}_i(k|k-1) )$ accounts for the lost packet.  
\end{rem}

For the following analysis, we assume bounded disturbances:
\begin{assum}
\label{ass:bounded_di}
For all $i \in \N_N$, $\|d_i \|_\infty \leq d_i^\text{\normalfont{max}}$.
\end{assum}
This assumption is realistic, when $d_i$ represents imperfect initialization or imprecise computations, for example.
Even though the assumption may not hold for the case of modeling packet drops in general, the developed method was found to be effective also for this scenario in the simulation examples of \sect \ref{sec:illustrEx}. 
%

With \eqref{eq:inputEstSingle}
and \eqref{eq:EBSE2}, the error \eqref{eq:epsilon_ci} becomes 
\begin{align}
&e_i(k) 
=(I-LC)A e_i(k-1) +(I-LC) B \tilde{u}(k-1)  \nonumber \\
&\phantom{=}+\! \sum\nolimits_{\ell \in \bar{I}(k)} \!\! L_\ell \big( y_\ell(k) - C_\ell \hat{x}_i(k|k-1) \big) -d_i(k) .
\label{eq:epsilon_ci_di_preStep}
\end{align}
The key difficulty that arises for the stability analysis is that the event-triggers \eqref{eq:eventTrigger_MB} bound the difference $\|y_i(k) - C_i \hat{x}_i(k|k-1) \|$ for agent $i$, but not the difference $\| y_\ell(k) - C_\ell \hat{x}_i(k|k-1) \|$ for $\ell \not= i$ as appearing in \eqref{eq:epsilon_ci_di_preStep}.
Hence, 
it can no longer be directly concluded that the sum in \eqref{eq:epsilon_ci_di_preStep} is bounded.
To obtain further insight, we rewrite \eqref{eq:epsilon_ci_di_preStep} as
\begin{align}
e_i(k) 
&=(I-LC)A e_i(k-1) +(I-LC) B \tilde{u}(k-1)  \nonumber \\
&\phantom{=}+\!\! \sum\nolimits_{\ell \in \bar{I}(k)} \!\! L_\ell \big( y_\ell(k) - C_\ell \hat{x}_\ell(k|k-1) \big) -d_i(k) \nonumber \\
&\phantom{=}-\!\! \sum\nolimits_{j \in \bar{I}(k)} \!\! L_j C_j A e_{ij}(k-1)
\label{eq:epsilon_ci_di}
\end{align}
where $e_{ij}(k) = \hat{x}_i(k) - \hat{x}_j(k)$ is the inter-agent error; we used $\hat{x}_i(k|k-1) - \hat{x}_\ell(k|k-1) = A e_{i \ell}(k-1)$; and we substituted index $j$ for $\ell$ in the last term.
In contrast to the simplified analysis in \sect \ref{sec:DEBSE_idealized}, agent $i$'s estimation error $e_i$ now depends on the other estimates through $e_{ij}$.
Except for the term $\sum_{j \in \bar{I}(k)} L_j C_j A e_{ij}(k-1)$, all other input terms are bounded by \eqref{eq:Ibar}, \Assump \ref{ass:bounded_di}, and \Lem \ref{lem:inputErrorsBounded}.  
Thus, for establishing stability, the inter-agent error $e_{ij}(k-1)$ must be bounded.  

The inter-agent error dynamics are given by
\begin{align}
e_{ij}(k) 
&= \hat{x}_i(k) - \hat{x}_j(k) 
=A e_{ij}(k-1) 
\nonumber \\
&\phantom{=} + \sum\nolimits_{\ell \in I(k)} L_\ell \big( y_\ell(k) - C_\ell \hat{x}_i(k|k-1) \big) +d_i(k) \nonumber \\
&\phantom{=} - \sum\nolimits_{\ell \in I(k)} L_\ell \big( y_\ell(k) - C_\ell \hat{x}_j(k|k-1) \big) -d_j(k) \nonumber \\
&= 
\tilde{A}_{I(k)} \, e_{ij}(k-1)
+d_i(k) -d_j(k) \label{eq:epsij_dyn}
\end{align}
where $\tilde{A}_{I(k)} := (I - \sum_{\ell \in I(k)} L_\ell  C_\ell)A$ 
was introduced for brevity.
The inter-agent error $e_{ij}(k)$ is governed by the unforced time-varying dynamics $e_{ij}(k) = \tilde{A}(k) e_{ij}(k-1)$. In general, one cannot infer stability of the event-based scheme from stability of the centralized design (\Assump \ref{ass:centralEst}) as was done in \sect \ref{sec:DEBSE_idealized} by disregarding inter-agent errors.
In fact, an example is presented in 
\ifextended
\sect \ref{sec:illustrExBC} 
\fi
where the inter-agent differences destabilize the event-based system despite stability of the centralized estimator.
If, however, the time-varying dynamics $e_{ij}(k) = \tilde{A}_{I(k)} e_{ij}(k-1)$ can be shown to be stable for a specific design $L$, boundedness of $e_{ij}(k)$ can be concluded.  Such an example is presented in
\ifextended
\sect \ref{sec:illustrExThermofluid}.
\fi

Next, we present a straightforward extension of the event-based communication scheme, which guarantees stability for an arbitrary centralized observer gain $L$.

\subsection{Synchronous averaging mechanism}
\label{sec:syncAvg}
Since the inter-agent error $e_{ij}(k)$ is the difference between the state estimates by agent $i$ and $j$, we have full control over it: we can make it zero at any time by resetting the two agents' state estimates to the same value, for example, their average. 
Therefore, a straightforward way to guarantee 
bounded inter-agent errors is to periodically reset all agents' estimates to their joint average.
Clearly, this strategy increases the communication load on the network.  If, however, the disturbances $d_i$ are small or only occur rarely, the required resetting period can be large relative to the underlying sampling time $T_s$.

We assume that the resetting happens after all agents have made their estimator updates \eqref{eq:EBSE2}.
Let $\hat{x}_i(k-) = \hat{x}_i(k)$ and $\hat{x}_i(k+)$ denote agent $i$'s estimate before and after resetting, and let $K \in \N$ be the fixed resetting period.  Each agent $i$ implements the following synchronous averaging mechanism: 
\begin{align}
&\text{If $k$ a multiple of $K$:} 
\!\!\!&& \text{transmit $\hat{x}_i(k-)$;} \label{eq:syncAvg} \\
& && \text{receive $\hat{x}_j(k-), j \in \N_N \!\setminus \!\{i \}$;} \nonumber \\
& && \text{set $\hat{x}_i(k+) = \avg\nolimits_{j=1,\dots,N} (\hat{x}_j(k-))$} \nonumber
\end{align}
%
We assume that the network capacity is such that the mutual exchange of the estimates can happen in one time step (as is the case for the system in \cite{Tr12}), and that no data is lost in the transfer (\eg through appropriate low-level protocols using acknowledgments).  
%
In other scenarios, one could take several time steps to exchange all estimates, at the expense of a delayed reset. Alternatives to resetting the average are also conceivable, such as resetting to the estimate of one particular agent.  
%

The synchronous averaging period $K$ can be chosen from simulations assuming a model for the inter-agent disturbances $d_i$ (\eg packet drops).  In a typical design procedure, an event-based design is carried out first making the idealizing assumption of zero inter-agent errors (according to \sect \ref{sec:DEBSE_idealized}). The triggering thresholds $\deltae_i$ are adjusted such that a desirable trade-off between communication and estimation performance is obtained.  In a second stage, \eqref{eq:syncAvg}
is introduced, and $K$ is chosen so as to keep inter-agent errors small.

\begin{rem}
The synchronous averaging \eqref{eq:syncAvg} requires exchange of estimates between agents, which is in addition to the event-based communications \eqref{eq:eventTrigger_MB} and \eqref{eq:eventTriggerCtrl}. Even though a stable estimation scheme with the synchronous exchange of estimates alone would be conceivable, this would typically require relatively high update rates. The design herein primarily relies on event-based communication.  The synchronous averaging mechanism is introduced just to keep inter-agent errors small and guarantee stability under practical circumstances.  In a scenario, where inter-agent differences occur infrequently, the synchronous resetting frequency can be small and most communication is due to the event-triggering.  
\end{rem}

\subsection{Stability result}
With the augmentation of the previous subsection, we can now establish a bounded difference between the EBSE and the centralized reference estimator.
\begin{thm}
\label{thm:epsilon_ci_notIdealized}
Under Assumptions \ref{ass:centralEst} and \ref{ass:bounded_di}, the difference $e_i(k)$ between the centralized estimator and the EBSE given by \eqref{eq:EBSE1}, \eqref{eq:inputEstSingle}, \eqref{eq:EBSE2}, \eqref{eq:eventTrigger_MB}, \eqref{eq:eventTriggerCtrl}, and \eqref{eq:syncAvg}  is bounded for 
all initial conditions $\hat{x}_i(0)$ and $x_\text{\normalfont{c}}(0)$.
\end{thm}
\begin{pf}
Since the agent error dynamics \eqref{eq:epsilon_ci_di} will be affected by the resetting \eqref{eq:syncAvg}, we first rewrite the error dynamics in terms of the average estimate $\bar{x}(k) := \avg( \hat{x}_i(k) )$.  Defining the errors $\bar{e}(k) := \hat{x}_\text{c}(k) - \bar{x}(k)$ and $\bar{e}_i(k) := \bar{x}(k) - \hat{x}_i(k)$, we have $e_i(k) = \bar{e}(k) + \bar{e}_i(k)$ and will establish the claim by showing boundedness of $\bar{e}_i(k)$ and $\bar{e}(k)$ separately.

For the average estimate $\bar{x}(k)$,
we obtain from \eqref{eq:EBSE1}, \eqref{eq:EBSE2},
\begin{align*}
&\bar{x}(k|k-1) = A \bar{x}(k-1) + B  \hat{u}(k-1) \\
&\bar{x}(k) = \bar{x}(k|k\!-\!1) + \!\!\! \sum_{\ell \in I(k)} \!\! L_\ell \big( y_\ell(k) - C_\ell \bar{x}(k|k\!-\!1) \big) + \bar{d}(k)
\end{align*}
where $\bar{x}(k|k-1) := \avg(\hat{x}_i(k|k-1))$
and $\bar{d}(k) := \avg(d_i(k))$.
%
%
The dynamics of the error $\bar{e}_i(k)$ are described by 
\begin{align}
\bar{e}_i(k) 
&= 
\tilde{A}_{I(k)} \bar{e}_i(k-1) 
+\bar{d}(k)-d_i(k) 
 \label{eq:epsj}  \\
\bar{e}_i(k+) &= 0, \qquad \text{for $k = \kappa K$ with some $\kappa \in \N$} \label{eq:epsj_reset} 
\end{align}
where \eqref{eq:epsj} is obtained by direct calculation analogous to \eqref{eq:epsij_dyn}, and \eqref{eq:epsj_reset} follows from \eqref{eq:syncAvg}.  That is, $\bar{e}_i$ is periodically reset to $0$ and evolves according to \eqref{eq:epsj} in-between resetting instants. 
Since $\bar{d}(k)$, $d_i(k)$, 
and $\tilde{A}_{I(k)}$ are all bounded, boundedness of  $\bar{e}_i$ for all $i \in \N_N$ follows.

Since $\bar{e}(k) = \avg_i(e_i(k))$, we obtain from \eqref{eq:epsilon_ci_di}
\begin{align}
\bar{e}(k)
&= (I-LC)A \bar{e}(k-1)   +(I-LC) B \tilde{u}(k-1)  \nonumber \\
&\phantom{=}+\!\! \sum\nolimits_{\ell \in \bar{I}(k)} \!\! L_\ell \big( y_\ell(k) - C_\ell \hat{x}_\ell(k|k-1) \big) -\bar{d}(k) \nonumber \\
&\phantom{=}-\!\! \sum\nolimits_{j \in \bar{I}(k)} \!\! L_j C_j A \bar{e}_{j}(k-1) 
\label{eq:eavg}
\end{align}
where 
we used $\avg_i(e_{ij}(k)) = \avg_i (\hat{x}_i(k) - \hat{x}_j(k)) = \bar{x}(k) - \hat{x}_j(k) = \bar{e}_j(k)$.
Note that \eqref{eq:eavg} fully describes the evolution of $\bar{e}(k)$.  In particular, the resetting \eqref{eq:syncAvg} does not affect $\bar{e}(k)$ because, at a resetting instant $k=\kappa K$, it holds that
$\bar{e}(k+) = \hat{x}_\text{c}(k) - \frac{1}{N} \sum\nolimits_{j=1}^N \hat{x}_j(k+) =\hat{x}_\text{c}(k) - \frac{1}{N} \sum\nolimits_{j=1}^N \big( \frac{1}{N} \sum\nolimits_{\ell=1}^N \hat{x}_\ell(k-) \big) =\hat{x}_\text{c}(k) - \frac{1}{N} \sum\nolimits_{\ell=1}^N \hat{x}_\ell(k-) = \bar{e}(k-)$.
All input terms in \eqref{eq:eavg} are bounded: $\tilde{u}$ 
by \Lem \ref{lem:inputErrorsBounded}, $\bar{d}$ by \Assump \ref{ass:bounded_di}, $\sum_{\ell \in \bar{I}(k)} L_\ell (y_\ell(k) - C_\ell \hat{x}_\ell(k|k-1))$ by the event-triggering mechanism (see \eqref{eq:Ibar}), and $\bar{e}_j$ for all $j$ by the previous argument.  Since $z(k) = (I-LC)A z(k-1)$ is exponentially stable, it follows that $\bar{e}(k)$ is bounded for all $k$ \cite[Thm.~75, p.~218]{CaDe91}, which completes the proof.
\hfill \ \qed
\end{pf}

By means of \The \ref{thm:epsilon_ci_notIdealized} and \eqref{eq:estError_ei}, properties about the agent's estimation error $\epsilon_i(k) = x(k) - \hat{x}_i(k)$ can readily be derived given properties of the disturbances $v$, $w$, and the centralized estimator.  The discussion in \sect \ref{sec:idealStabAnalysis} applies analogously.
%
An upper bound on $\epsilon_i(k)$ in terms of the problem parameters corresponding to \eqref{eq:thm_epsilon_ci_idealized} can be derived for \The \ref{thm:epsilon_ci_notIdealized}; it is omitted here for simplicity.  If the inter-agent errors are small, the simpler bound \eqref{eq:thm_epsilon_ci_idealized} may be used for all practical purposes.  
%

\section{Distributed Event-based Control}
\label{sec:DEBcontrol}
In this section, we discuss stability of the full distributed, event-based control system shown in \fig \ref{fig:oneAgent}, where the estimate $\hat{x}_i$ is used for control.
Algorithm \ref{alg:EBimplementation} summarizes the implementation of the complete event-based control system on each agent.
%
%
\begin{algorithm}[ht]
\caption{{\small Implementation on agent $i$, executed every time step $k$.}}
%
\label{alg:EBimplementation}
\scriptsize 
\vspace{0.5ex}
\begin{tabbing}
\hspace*{1ex}\=\hspace*{4ex}\=\hspace*{4ex}\=\hspace*{4ex}\= \kill
\>Actuation: apply $u_i(k-1)$\\ 
\>Sensing: acquire $y_i(k)$\\
\>Estimation:\\
\>\>Prediction \eqref{eq:EBSE1}: compute $\hat{x}_i(k|k-1)$\\
\>\>Measurement trigger \eqref{eq:eventTrigger_MB}: send/receive $y_\ell(k), \ell \in I(k)$\\
\>\>Measurement update \eqref{eq:EBSE2ideal}: compute $\hat{x}_i(k|k) = \hat{x}_i(k)$ \\
\>\>{\bf if} ($k$ is a multiple of $K$) \\
\>\>\>perform synchronous reset \eqref{eq:syncAvg} \\
\>\>{\bf end if} \\
\>Control:\\
\>\>Control law \eqref{eq:stateFeedbackDistributed}: compute $u_i(k)$ from $\hat{x}_i(k)$\\
\>\>Control trigger \eqref{eq:eventTriggerCtrl}: send/receive $u_\ell(k), \ell \in I^\text{ctrl}(k)$
\end{tabbing}
\vspace{0.5ex}
\end{algorithm}

Closed-loop stability can be deduced from stability of the event-based estimation (\The \ref{thm:epsilon_ci_notIdealized}) and of the centralized control system (\Assump \ref{ass:centralCtrl}).  
%
%
\begin{thm}
\label{thm:boundednessControl}
Let $v$ and $w$ be bounded, and let assumptions \ref{ass:centralEst}, \ref{ass:centralCtrl}, \ref{ass:bounded_di} be satisfied.
Then, the state $(x(k)$, $\epsilon_1(k)$, $\epsilon_2(k), \dots, \epsilon_N(k))$ of the distributed event-based control system given by \eqref{eq:system_x}, \eqref{eq:system_y}, \eqref{eq:stateFeedbackDistributed}, \eqref{eq:EBSE1},  \eqref{eq:inputEstSingle}, \eqref{eq:EBSE2}, \eqref{eq:eventTrigger_MB}, \eqref{eq:eventTriggerCtrl}, \eqref{eq:syncAvg}
is bounded for all initial conditions $\hat{x}_i(0)$ and $x(0)$.
\end{thm}
\begin{pf}
From \Assump \ref{ass:centralEst} with bounded $v$ and $w$, it follows that the estimation error $\epsilon_\text{c}(k)$ of the centralized observer is bounded (\cf \eqref{eq:closedLoopCentralized_est}).  Thus, \eqref{eq:estError_ei} and \The \ref{thm:epsilon_ci_notIdealized} imply that all  $\epsilon_i$, $i \in \N_N$, are bounded. 
Using \eqref{eq:stateFeedbackDistributed} and \eqref{eq:estError_ei}, 
we rewrite \eqref{eq:system_x_long} as
$x(k) = (A+BF) x(k\!-\!1) - \! \sum_{i \in \N_N} B_i F_i \epsilon_i(k\!-\!1) + v(k\!-\!1)$.
 Hence, it follows from stability of $A+BF$ (\Assump \ref{ass:centralCtrl}) and bounded $v$ that $x$ is also bounded.
\hfill \ \qed
\end{pf}

\ifextended
\section{Illustrative Examples}
\fi
\label{sec:illustrEx}

\ifextended
The proposed methods for distributed event-based estimation and control are illustrated by means of two simulations examples: a simulation model of the Balancing Cube test bed \citep{TrDAn12b}, 
which was also used for the experiments in \citep{Tr12}, as well as a thermo-fluid process benchmark problem \citep{GrHiJuEtAl14}.
%
\fi
Matlab files to run both simulation examples are provided as supplementary material.\footnote{Available at  
\url{http://is.tue.mpg.de/publications/tr17} 
or from the author's web page.}

\ifextended
\subsection{Balancing Cube}
\label{sec:illustrExBC}
\fi

The Balancing Cube is a dynamic sculpture that can balance on any one of its edges or corners by means of six rotating arms attached to its structure.  The arms constitute the agents of the NCS as in \fig \ref{fig:networkArchitecture}.
The system was developed as a test bed for distributed estimation and control and used, among others, to demonstrate the event-based algorithms herein, \citep{Tr12}.



\ifextended
\subsubsection{System description}
\fi
A state-space model \eqref{eq:system_x} of the cube's linearized dynamics when balancing on an edge is given in \citep[Eq.~(3)]{TrDAn12c}.  Analogous to the implementation in \citep{Tr12}, we exploit the model structure to obtain a reduced-state model for estimator design:
\begin{align}
&x(k) = A x(k\!-\!1) + B_1 u(k\!-\!1) + B_2 u(k\!-\!2) + v(k\!-\!1) 
\nonumber \\
&\quad A =
\begin{bmatrix}
I &  0 \\[\spaceMatRow]
A_{31} & A_{33}
\end{bmatrix}\! , \,
B_1 = 
\begin{bmatrix}
T_\text{s} I \\[\spaceMatRow]
B_3
\end{bmatrix}\! , \,
B_2 = 
\begin{bmatrix}
0 \\[\spaceMatRow]
 A_{32}
\end{bmatrix} \label{eq:BC_model_x_red}
\end{align}
with sampling time $T_\text{s} = 0.01 \, \text{s}$, $I$ and $0$ identity and zero-matrices, and the numerical values of the other matrices given in \citep{TrDAn12c}.  The model includes the effect of local velocity feedback on each agent, and the inputs $u$ are the reference velocities to these local controllers.  The special structure of this model 
follows from a time-scale separation algorithm assuming sufficiently fast velocity feedback loops
(see \citep{TrDAn12b} for details).  
The states and inputs of the model \eqref{eq:BC_model_x_red} are summarized in Table~\ref{tab:StatesAndInputsBC}.

%

\begin{table}
\renewcommand{\arraystretch}{1.1}
\caption{States and inputs of the Balancing Cube model.
}
\label{tab:StatesAndInputsBC}
\centering
\scriptsize
\begin{tabular}{lll}
\hline 
{\bf States} && {\bf Unit}  \\ 
\hline 
$x_1, \dots, x_6$ & angle arm 1 to 6  & $\text{rad}$ \\
$x_{7}$ & angle cube body  & $\text{rad}$ \\
$x_{8}$ & angular velocity cube body  & $\text{rad}/\text{s}$ \\
\hline
{\bf Inputs} && {\bf Unit}  \\ 
\hline 
$u_1, \dots, u_6$ & reference angular velocity & $\text{rad}/\text{s}$ \\
\hline 
\end{tabular} 
%
\end{table}

%

The output equation \eqref{eq:system_y} is
%
\begin{align}
y(k) &= 
\begin{bmatrix}
y_1(k) \\[\spaceMatRow]
\vdots \\[\spaceMatRow]
y_9(k)
\end{bmatrix}
=
\left[
\begin{smallmatrix}
1 & 0 & 0 & 0 & 0 & 0 & 0 & 0 \\
0 & 0 & 0 & 0 & 0 & 0 & 0 & 1 \\
0 & 1 & 0 & 0 & 0 & 0 & 0 & 0 \\
0 & 0 & 0 & 0 & 0 & 0 & 0 & 1 \\
0 & 0 & 1 & 0 & 0 & 0 & 0 & 0 \\
0 & 0 & 0 & 0 & 0 & 0 & 0 & 1 \\
0 & 0 & 0 & 1 & 0 & 0 & 0 & 0 \\
0 & 0 & 0 & 0 & 1 & 0 & 0 & 0 \\
0 & 0 & 0 & 0 & 0 & 1 & 0 & 0 \\
\end{smallmatrix}
\right]
x(k)
+ w(k) 
\label{eq:BC_model_y}
\end{align}
which is in accordance with the sensors on the physical system 
(angle encoders for the six arms $y_1$, $y_3$, $y_5$, $y_7$, $y_8$, $y_9$, and rate gyroscopes $y_2$, $y_4$, $y_6$).
Inputs and outputs as associated to the six agents are summarized in \tab \ref{tab:AgentsBC}.
As is easily checked (see supplementary files), the state $x$ is not detectable from the sensors of any individual agent. Inter-agent communication is therefore necessary for stable state estimation.
\begin{table}
\renewcommand{\arraystretch}{1.1}
\centering
\caption{Agents of the Balancing Cube model.
%
}  
\label{tab:AgentsBC}
\scriptsize
%
%
\begin{tabular}{|l|llllll|}
\hline 
{\bf Agent \#} & 1 & 2 & 3 & 4 & 5 & 6 \\ \hline 
{\bf Actuator}  & $u_1$ & $u_2$ & $u_3$ & $u_4$ & $u_5$ & $u_6$ \\ \hline
{\bf Sensors}  & $y_1$, $y_2$ & $y_3$, $y_4$ & $y_5$, $y_6$ & $y_7$ & $y_8$ & $y_9$ \\ \hline 
\end{tabular} 
\end{table}


The noise variables $v(k)$ and $w(k)$ are modeled as independent uniform random variables.  The sensor noise variance is chosen comparable to the sensors on the actual system (noise on angle measurements is negligible, noise on angular rate sensors is significant).  To account for non-ideal actuation, we simulate input noise, uniformly distributed in $[-0.05, 0.05] \, \text{rad/s}$.  For any measurement $y_i(k)$ transmitted over the network, a packet drop probability of 2\% is modeled. Dropped measurements can be expressed by the disturbance signal $d_i$ as explained in \Rema \ref{rem:packetDrops}.
For simplicity, we assume that inputs and synchronous resetting are
not affected by data loss.

\ifextended
\subsubsection{Event-based estimation and control system design}
\fi
\label{sec:BC_EBdesign}
Each agent implements \algo \ref{alg:EBimplementation}, but in slight deviation thereof, 
agents 1 to 3 make individual triggering decisions for each of their two sensors.
For example, Agent 2 implements the trigger \eqref{eq:eventTrigger_MB} separately for $y_3$ and $y_4$.  Advantages are greater potential for reducing the amount of transmitted data and different units are handled naturally. In other scenarios such as networks allowing for large packet sizes, sending all measurements at once may be  preferred.
The proposed framework can accommodate for these different communication architectures.

To implement the event-based estimation and control scheme, observer and controller gains, and the communication thresholds need to be chosen.  The centralized observer \eqref{eq:FCSE1}, \eqref{eq:FCSE2} is designed as a steady-state Kalman filter with eigenvalues of the error dynamics $(I-LC)A$ at $0.97$, $0.96$, and $0.73$ (sixfold). 
The centralized controller \eqref{eq:stateFeedback} is obtained from 
an LQR design guaranteeing stability of $A+BF$ (see supplementary material for details).
We chose the following triggering thresholds: $\deltae_\text{gyro} = 0.004 \, \text{rad}/\text{s}$ for the rate gyro sensors ($y_2$, $y_4$, $y_6$), $\deltae_\text{enc} = 0.001 \, \text{rad}$ for the encoders (all other $y_i$), and $\deltac = 0.01 \, \text{rad}/\text{s}$ for all inputs. 
In this example, we shall see that synchronous averaging \eqref{eq:syncAvg} is required for stability.  We chose $K=200$ as the reset period (every $2 \, \text{s}$).

\begin{figure}[tb]
\centering
\includegraphics{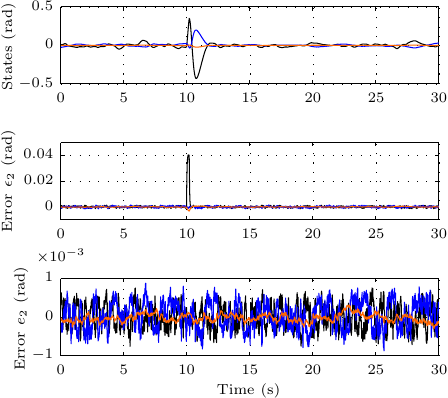}
\caption{Simulation of the Balancing Cube model.  From top to bottom: states $x$, estimation errors $\epsilon_2$ of agent 2, and its difference $e_2$ to a centralized estimate.  In each graph, three signals are shown: the angles of arm 2 ({\bf black}), arm 5 ({\bf blue}), and the cube structure ({\bf orange}).  
}
\label{fig:exampleSimRum_BC_x}
\end{figure}

%
\ifextended
\subsubsection{Simulation results}
\fi
Figure \ref{fig:exampleSimRum_BC_x} shows selected state and error signals from simulating the event-based control system
as described in \sect \ref{sec:BC_EBdesign}.
The error signals are given for agent 2; both the actual estimation error $\epsilon_2$ and the difference $e_2$ to the centralized reference estimator are shown.
The difference $e_2$ to the centralized estimate is guaranteed to be bounded by \The \ref{thm:epsilon_ci_notIdealized}.  The results show that the effective bounds on the arm angles (black and blue graphs) are in the order of
the triggering threshold $\deltae_\text{enc} = 0.001 \, \text{rad}$.
In general, however, the bounds depend not only on the trigger thresholds, but also the error dynamics as per
\eqref{eq:epsilon_ci_di}.

\begin{figure}[tb]
\centering
\includegraphics{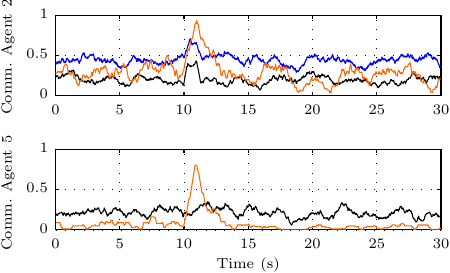}
\caption{Event-based communication rates for two agents in the cube example. {\bf Black:} encoder measurements; {\bf blue:} rate gyro measurements (Agent 5 has no such sensor); {\bf orange:} control inputs.  All rates as moving average over 100 steps.}
\label{fig:exampleSimRum_BC_comm}
\end{figure}

At $10 \, \text{s}$, an impulsive disturbance was applied at the input of agent 2. 
From the corresponding communication rates given in \fig \ref{fig:exampleSimRum_BC_comm}, a temporary increase in communication can be observed in response to the disturbance. 
Also visible in \fig \ref{fig:exampleSimRum_BC_comm} is that encoder and input communication rates are generally lower than those for the rate gyro sensors.  This can be expected as the rate gyros are responsible for observing the unstable mode of the system. 
In general, the event-based design is effective in adapting communication: 
when there is need for feedback (\eg disturbances or unstable modes), rates go up.  

\begin{figure}[tb]
\centering
\includegraphics{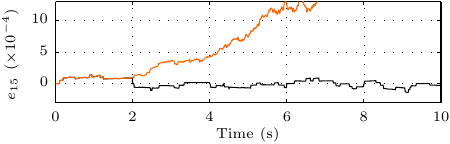}
\caption{Inter-agent error $e_{15}$ between agents 1 and 5 (one component). {\bf Black:} with synchronous averaging; {\bf orange:} without.
}
\label{fig:exampleSimRum_BC_interAgentError}
\end{figure}

Figure \ref{fig:exampleSimRum_BC_interAgentError} shows one component of the inter-agent error $e_{15}$; that is, the difference between the estimates of agent 1 and 5.  The black graph corresponds to a short sequence of the same simulation run as in \fig \ref{fig:exampleSimRum_BC_x} and \ref{fig:exampleSimRum_BC_comm}.  
Synchronous averaging \eqref{eq:syncAvg} with $K=200$ causes all inter-agent errors to reset to zero every $2 \, \text{s}$.
In contrast, the orange graph corresponds to a simulation run with \eqref{eq:syncAvg} disabled ($K \to \infty$).  Obviously, the inter-agent error drifts meaning the estimates of the two agents slowly deviate, which eventually destabilizes the whole closed-loop system.  
With synchronous resetting, however, the inter-agent errors $e_{ij}$ are an order of magnitude smaller than the estimation errors $e_i$ or $\epsilon_i$ (\cf \fig \ref{fig:exampleSimRum_BC_x} and \ref{fig:exampleSimRum_BC_interAgentError}).  Thus, one can use the simplification of \sect \ref{sec:DEBSE_idealized} for the purpose of analyzing the performance of the system.  Yet, for stability, considering non-zero inter-agent errors is critical as this example highlights.
\ifextended
In contrast to this example, the thermo-fluid example in \sect \ref{sec:illustrExThermofluid} is one where synchronous resetting is found to be unnecessary.
\fi

\begin{figure}[tb]
\centering
\includegraphics{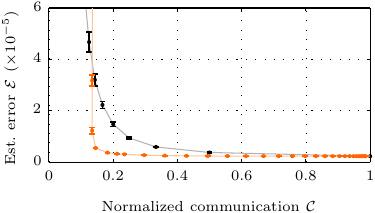}
\caption{Estimation error versus communication trade-off for event-based ({\bf orange}) and periodic communication ({\bf black}). The error bars represent plus/minus one standard deviation of the obtained values for $\mathcal{E}$ from 100 simulation runs. The variability in $\mathcal{C}$ is comparably small and thus not shown.
}
\label{fig:exampleSimRum_BC_EvsC}
\end{figure}

\ifextended
\subsubsection{Estimation versus communication trade-off}
\fi
The event-based estimation scheme herein allows the designer to trade off estimation performance for communication by choosing the triggering thresholds. We illustrate this trade-off by performing simulations for different choices of the thresholds.  For this, the previously chosen thresholds (\sect \ref{sec:BC_EBdesign})
are multiplied with a factor varying from 0 to 3 on a logarithmic scale.  For each parameter setting, we perform 100 independent simulation runs and compute normalized measures for estimation and communication.
The normalized estimation error $\mathcal{E}$ is defined as the squared error $\epsilon_i^\transpose\!(k) \epsilon_i(k)$, averaged over simulation horizon, all agents, and simulation runs. 
The normalized communication $\mathcal{C}$ is obtained by averaging the number of transmitted scalars (measurements, inputs, resets) such that $\mathcal{C}=1$ corresponds to full periodic communication of all measurements and inputs, and $\mathcal{C}=0$ to no communication.
%
%
The obtained $\mathcal{E}$-vs-$\mathcal{C}$ graph is shown in \fig \ref{fig:exampleSimRum_BC_EvsC} for the event-based communication design (orange), in comparison to periodic communication (black).
%
%
The periodic communication graph corresponds to communicating inputs and measurements at multiples of the discrete sampling time $T_\text{s}$ (i.e., $1 T_\text{s}$, $2 T_\text{s}$, $3 T_\text{s}$, etc).  Controller and observer gains are recomputed for the re-sampled discrete-time model.

%

While for full communication $\mathcal{C} = 1$ both designs yield the same error as expected, event-based communication is superior for lower communications; roughly 80\% communication can be saved before experiencing any significant increase in the estimation error.
Yet, for very low communication, slow periodic sampling is superior in this example.  In general, the trade-off is multi-dimensional due to multiple triggering thresholds, while \fig \ref{fig:exampleSimRum_BC_EvsC} only represents one dimension.  Nonetheless, this example shows the potential of event-based sampling to make better average use of available communication resources than standard periodic sampling.
%


\ifextended 	

\subsection{Thermo-fluid process}
\label{sec:illustrExThermofluid}
As a second example, we consider distributed event-based control of a thermo-fluid process.  In contrast to the previous example, no synchronous resetting is necessary here.  The process has been proposed as a benchmark problem for event-based control by \cite{GrHiJuEtAl14}, and used to demonstrate different event-based methods in \citep{StVeLu13,SiStGrLu15}, for example.
%

The key components of the plant are two tanks containing fluids, whose level and temperature are to be regulated by controlling the tanks' inflows, as well as heating and cooling units.  Both tanks are subject to step-like disturbances, and their dynamics are coupled through cross-flows between the tanks.  Each tank is associated with a control agent responsible for computing commands to the respective actuators.  Each agent can sense the temperature and level of its tank. For details on the physical process, including diagrams and photos, the reader is referred to \citep{GrHiJuEtAl14,SiStGrLu15}.
%


\subsubsection{System description}
\label{sec:SystemDescriptionThermofluid}
The discrete-time linear model \eqref{eq:system_x}, \eqref{eq:system_y} representing the thermo-fluid process dynamics about an equilibrium configuration is obtained by zero-order hold discretization  with a sampling time $T_\text{s} = 0.2 \, \text{s}$ of the continuous-time model given in \cite[Sec.~5.8]{GrHiJuEtAl14}.
In contrast to the Balancing Cube example, the process dynamics are stable.
The states and inputs of the system are summarized in Table~\ref{tab:StatesAndInputsThermofluid}.
Same as \cite{SiStGrLu15}, we consider step-wise process disturbances $v(k)$ from exogenous inflow and heating.  
%
Noisy measurements $y(k) \in \R^4$ of all states are available (\ie $C=I$ in \eqref{eq:system_y}), and sensor noise $w(k)$ is modeled as uniformly distributed random variable.
The numerical parameters of the model are available in the supplementary files.

\begin{table}
\renewcommand{\arraystretch}{1.1}
\caption{States and inputs of the thermo-fluid process.
}
\label{tab:StatesAndInputsThermofluid}
\centering
\scriptsize
\begin{tabular}{lll}
\hline 
{\bf States} && {\bf Unit}   \\ 
\hline 
$x_1(k)$ & level tank 1  & m  \\
$x_2(k)$ & temperature tank 1 & K   \\
$x_3(k)$ & level tank 2  & m  \\
$x_4(k)$ & temperature tank 2 & K   \\ 
\hline 
{\bf Inputs} && {\bf Unit}   \\ 
\hline 
$u_{11}(k)$ & inflow tank 1  & 1 (normalized) \\
$u_{12}(k)$ & cooling tank 1  & 1 (normalized) \\
$u_{21}(k)$ & inflow tank 2  & 1 (normalized) \\
$u_{22}(k)$ & heating tank 2  & 1 (normalized) \\ 
\hline 
\end{tabular} 
\end{table}


Similar to the distributed architecture in \cite{SiStGrLu15}, we consider two agents (one for each tank) exchanging data with each other over a network link.  
Agent 1 measures $y_1$ (level) and $y_2$ (temperature) of its tank and is responsible for controlling $u_1 = (u_{11}, u_{12})$ (inflow and cooling); and Agent 2 accordingly.  
The agents'
associated sensors and actuators are summarized in \tab \ref{tab:AgentsThermofluid}.
%
\begin{table}[tb]
\renewcommand{\arraystretch}{1.1}
\caption{Agents in the thermo-fluid example. 
}  
\label{tab:AgentsThermofluid}
\centering
\scriptsize
%
%
\begin{tabular}{|l|ll|}
\hline 
{\bf Agent \#} & 1 & 2 \\ \hline 
{\bf Actuator}  & $u_1 = (u_{11}, u_{12})$ & $u_2 = (u_{21}, u_{22})$ \\ \hline
{\bf Sensors}  & $y_1$, $y_2$ & $y_3$, $y_4$  \\ \hline 
\end{tabular} 
\end{table}

In order to study the effect of imperfect communication, we simulate random packet drops.  Any measurement $y_i(k)$ that is transmitted between the two agents is lost with a probability of 5\%, independent of previous drops.  Packet drops are represented by the disturbance $d_i(k)$ in \eqref{eq:EBSE2} as discussed in \Rema \ref{rem:packetDrops}.
For simplicity, we assume that inputs when communicated are never lost.

\subsubsection{Event-based estimation and control system design}
Each agent implements \algo \ref{alg:EBimplementation}.  As in the example in \sect \ref{sec:illustrExBC}, triggering decisions are made individually for the two sensors of each agent (\cf \tab \ref{tab:AgentsThermofluid}).  However, to illustrate the flexibility of the proposed scheme, the triggering decision \eqref{eq:eventTriggerCtrl} is applied jointly for both inputs; for example, agent 2 makes the transmit decision jointly for $u_2 = (u_{21}, u_{22})$.


For the design of the centralized observer \eqref{eq:FCSE1}, \eqref{eq:FCSE2}, we chose $L =$ $\diag(0.1, 0.05, 0.1, 0.05)$ as observer gain, leading to stable centralized estimator dynamics $(I-LC)A$ with eigenvalues approximately at 0.95 and 0.9 (each one twice).
%
As it turns out, for this choice for $L$, also the switching inter-agent error dynamics \eqref{eq:epsij_dyn} are stable, which can be seen as follows.  By direct calculation, one can verify that
\begin{equation}
\tilde{A}_{J}^\transpose P \tilde{A}_{J} -P < 0
\quad \text{with} \,\, \tilde{A}_{J} := (I - \sum\nolimits_{\ell \in J} L_\ell  C_\ell)A
\end{equation}
is satisfied for $P = \diag(500, 1, 500, 1)$ and for all permutations $J \subseteq \{1,2,3,4\}$.  Thus, it follows that  $V(z) = z^\transpose P z$ is a common Lyapunov function (see \cite{MiFeMo00}) for the switching system $z(k) = \tilde{A}_{J(k)} z(k-1)$ for arbitrary $J(k) \subseteq \{1,2,3,4\}$, and, in particular, for $J(k) = I(k)$.  It thus follows that the unforced dynamics in \eqref{eq:epsij_dyn} are exponentially stable, which implies boundedness of $e_{ij}$.  Hence, synchronous averaging \eqref{eq:syncAvg} is not necessary in this example (in contrast to the one in \sect \ref{sec:illustrExBC}).

The state-feedback gain $F$ is obtained from a centralized LQR design \citep{AsWi97}.  In contrast to the decentralized design in \citep{SiStGrLu15}, the controller involves full couplings between all states.  Using a centralized control design with arbitrary couplings is made possible by maintaining an estimate of the full state on each agents in the propose architecture. 

Finally, the triggering thresholds in \eqref{eq:eventTrigger_MB} and \eqref{eq:eventTriggerCtrl} are set to $\deltae_1 = \deltae_3 = 0.01 \, \text{m}$, $\deltae_2 = \deltae_4 = 0.2 \, \text{K}$, and $\deltac_1 = \deltac_3 = 0.02$.  

\subsubsection{Simulation results}
The state trajectories of a $2000 \, \text{s}$ simulation run under event-based communication are shown in \fig \ref{fig:exampleSimRum_thermoFluid_x}. For comparison with the event-based estimates (orange), also the centralized estimate (blue) is shown.  
Step-wise disturbances $v$, active from $200 \, \text{s}$ to $600 \, \text{s}$ and from $1000 \, \text{s}$ to $1200 \, \text{s}$  (gray shaded areas) with comparable magnitudes as in \citep{SiStGrLu15}, cause the states to deviate from the equilibrium $x=0$.   Especially at times when disturbances are active, the event-based estimate is inferior to the centralized one as is expected due to the reduced number of measurements.  However, \The \ref{thm:epsilon_ci_notIdealized} guarantees that this difference is bounded and that it can be controlled by appropriate choice of the triggering thresholds.
%
\begin{figure}[tb]
\centering
\includegraphics{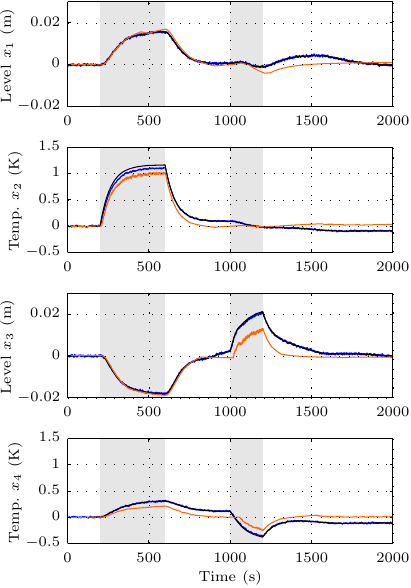}
\caption{State trajectories for the thermo-fluid simulation example.  {\bf Black:} the actual states $x$; {\bf orange:} event-based estimate $\hat{x}_1$ by Agent 1; {\bf blue:} centralized estimate $\hat{x}_\text{c}$.  The centralized estimate is shown for comparison and not available on any of the agents. The gray shaded areas indicate periods where step-wise process disturbances are active.}
\label{fig:exampleSimRum_thermoFluid_x}
\end{figure}

The average communication rates for event-based input and sensor transmissions 
 are given in \fig \ref{fig:exampleSimRum_thermoFluid_comm}.  Clearly, communication rates increase in the periods where the disturbances are active (gray areas), albeit not the same for all sensors and inputs.  At times, when there is no disturbance, communication rates are very low in general.  This highlights 
the intend behavior of event-based communication: sensor feedback only happens when necessary, in this case, when disturbances are active.
\begin{figure}[tb]
\centering
\includegraphics{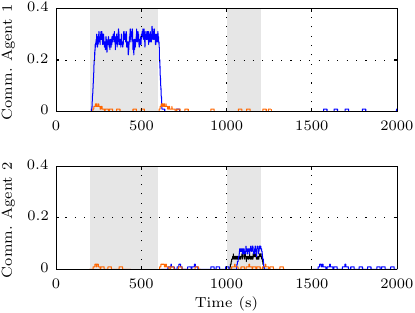}
\caption{Event-based communication rates for the thermo-fluid simulation example: communication of level measurements $y_1$ and $y_3$ in {\bf black}, for temperature measurements $y_2$ and $y_4$ in {\bf blue}, and for the inputs $u_1$ and $u_2$ in {\bf orange}.  All rates are computed as the moving average over 100 steps.}
\label{fig:exampleSimRum_thermoFluid_comm}
\end{figure}

Figure \ref{fig:exampleSimRum_thermoFluid_e12} shows the inter-agent error $e_{12}$; that is, the difference between the agents' estimates.  Jumps in the error signals are cause by a dropped packet.
Clearly, the error decays after each impulse, which corresponds to the exponentially stable inter-agent dynamics \eqref{eq:epsij_dyn} as per the previous discussion.  Because of this, no synchronous resetting \eqref{eq:syncAvg} is required in this example.
\begin{figure}[tb]
\centering
\includegraphics{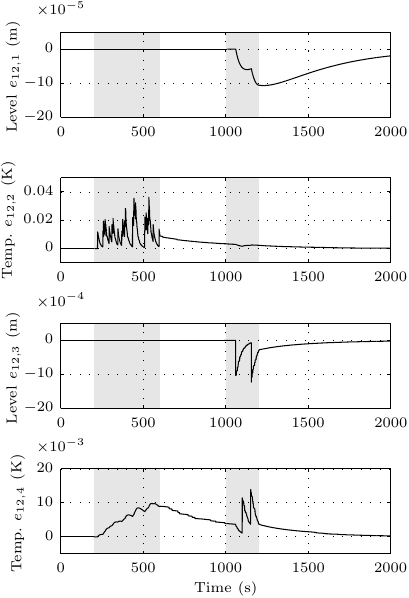}
\caption{Inter-agent error $e_{12}$ for the thermo-fluid example. Jumps in error are caused by packet drops, decay afterward is due to stable inter-agent dynamics \eqref{eq:epsij_dyn}.}
\label{fig:exampleSimRum_thermoFluid_e12}
\end{figure}

\fi

\section{Concluding Remarks}
\label{sec:conclusion}

Simplicity of design and implementation were key considerations when developing the approach for distributed and event-based state estimation herein.  The proposed scheme emulates a centralized linear state observer design, and directly builds upon the typical periodic implementation of an observer-based control system.  Starting from a centralized, periodic design (\sect \ref{sec:centralizedController}), only the event-triggers \eqref{eq:eventTrigger_MB} and \eqref{eq:eventTriggerCtrl}, and (for some problems) synchronous averaging \eqref{eq:syncAvg} must be added. Observer and controller structure, as well as the transmitted quantities (measurements and inputs) remain unchanged, and no redesign of gains is necessary.  The performance of the periodic design can be recovered by choosing small enough triggering thresholds, which greatly simplifies tuning in practice.  Thus, implementation of the event-based scheme requires minimal extra effort,
and virtually no additional design knowledge.

While conceptually straightforward, proving stability of the proposed architecture is non-trivial because the resulting observer dynamics are switching and coupled through process dynamics and communication. Together with the proposal of the architecture, establishing stability is the key contribution of this report. To prove stability for the general case, the synchronous averaging mechanism \eqref{eq:syncAvg} was introduced.  Yet, synchronous averaging can be avoided if the inter-agent dynamics \eqref{eq:epsij_dyn} are stable (as in the example in 
\ifextended
\sect \ref{sec:illustrExThermofluid}).
\fi
 Currently ongoing work concerns synthesis approaches, where the observer gains are not taken from a centralized design, but designed for stable inter-agent dynamics. 

The method proposed herein (albeit with full communication of inputs and without synchronous resetting), was successfully applied in \citep{Tr12} on the Balancing Cube test bed, i.e., a system with unstable dynamics.  
Comparable simulations of this system are presented herein.

\bibliographystyle{abbrvnat}        
{\small \bibliography{literature}}


\end{document}